\documentclass[aps,prl,twocolumn,superscriptaddress,english]{revtex4}
\usepackage{amssymb}
\usepackage{graphicx}
\usepackage{amsmath}
\usepackage{soul}
\usepackage{amsthm}
\usepackage{amsfonts}
\usepackage[T1]{fontenc}
\usepackage[latin9]{inputenc}
\usepackage{array}
\usepackage{multirow}
\usepackage{color}
\usepackage{esint}
\usepackage{bm}
\usepackage{color}
\usepackage{bbm}
\usepackage{hyperref}
\usepackage{babel}
\usepackage{titlesec}

\begin{document}

\global\long\def\id{\mathbbm{1}}
\global\long\def\ui{\mathbbm{i}}
\global\long\def\ud{\mathrm{d}}

\title{Dissipation-Driven Transition of Particles from Dispersive to Flat Bands}
\author{Yutao Hu}
\affiliation{Shenzhen Institute for Quantum Science and Engineering,
	Southern University of Science and Technology, Shenzhen 518055, China}
\affiliation{International Quantum Academy, Shenzhen 518048, China}
\affiliation{Guangdong Provincial Key Laboratory of Quantum Science and Engineering, Southern University of Science and Technology, Shenzhen 518055, China}
\author{Chao Yang}
\affiliation{Department of Physics, Southern University of Science and Technology, Shenzhen 518055, China}
\affiliation{Shenzhen Institute for Quantum Science and Engineering,
Southern University of Science and Technology, Shenzhen 518055, China}
\author{Yucheng Wang}
\thanks{Corresponding author: wangyc3@sustech.edu.cn}
\affiliation{Shenzhen Institute for Quantum Science and Engineering,
	Southern University of Science and Technology, Shenzhen 518055, China}
\affiliation{International Quantum Academy, Shenzhen 518048, China}
\affiliation{Guangdong Provincial Key Laboratory of Quantum Science and Engineering, Southern University of Science and Technology, Shenzhen 518055, China}
\begin{abstract}
Flat bands (FBs) play a crucial role in condensed matter physics, offering an ideal platform to study strong correlation effects and enabling applications in diffraction-free photonics and quantum devices. However, the study and application of FB properties are susceptible to interference from dispersive bands. Here, we explore the impact of a type of experimentally controllable dissipation on systems hosting both flat and dispersive bands by calculating the steady-state density matrix. We demonstrate that such dissipation can drive particles from dispersive bands into FBs and establish the general conditions for this phenomenon to occur. Our results demonstrate that dissipation can eliminate the influence of dispersive bands, thereby facilitating FB preparation, property measurement, and utilization. This opens a new avenue for exploring FB physics in open quantum systems, with potential implications for strongly correlated physics.
\end{abstract}
\maketitle

{\em Introduction.---} Flat band (FB) systems have attracted widespread attention due to their quenched kinetic energy, leading to eigenmodes that are compactly localized in space. This makes the system highly sensitive to interactions, often giving rise to strongly correlated phenomena such as fractional Chern insulators~\cite{Chern1,Chern2,Chern3}, superconductors~\cite{Chern3,SC1,SC2,SC3,SC4}, Mott insulators~\cite{MI1,MI2}, and others. Even without interactions, FB systems can exhibit many interesting phenomena, such as the inverse Anderson transition~\cite{AT1,AT2,AT3}, multifractal behavior and unconventional mobility edges~\cite{ME1,ME2,ME3}, influenced by disorder or quasi-periodic potentials. FBs have been realized in various platforms, including solid-state systems~\cite{exp1,exp2}, cavity polaritons~\cite{exp3,exp4,stub2}, photonic waveguides~\cite{exp5,exp6,exp7}, superconducting wire networks~\cite{SCwire1,SCwire2} and ultracold atoms~\cite{cold1,cold2,cold3,cold4,cold5,cold6,cold7,cold8}. In particular, recent advances in Moir\'{e} materials provide an intrinsic platform for studying the connection between FBs and quantum geometry, as well as the applications of FB systems~\cite{Chern3,SC1,SC2,Moire1,Moire2,Moire3,Moire4,Moire5}. 

\begin{figure}[t]
	\centering
	\includegraphics[width=\linewidth]{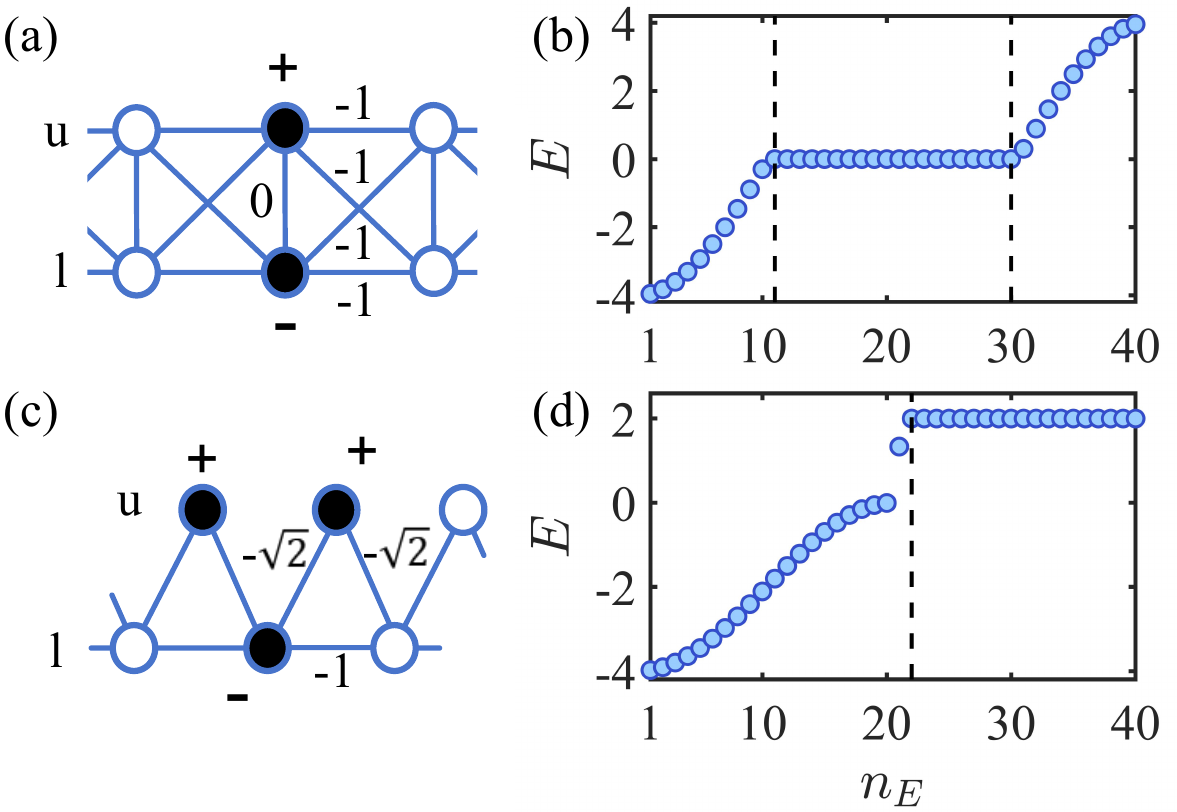}
	\caption{ Lattice structures and CLS configurations for the (a) cross-stitch ($U=1$) and (c) sawtooth ($U=2$) models. Filled circles denote sites occupied by the CLS.
		The corresponding energy spectra are shown in (b) and (d), respectively. For the cross-stitch model, the inter-cell and intra-cell hoppings are set to $-1$ and $0$, respectively. For the sawtooth model, the intra-cell hopping is $-\sqrt{2}$, with inter-cell hoppings of $-\sqrt{2}$ between upper and lower chains, and $-1$ between sites on the lower chain. Throughout this work, onsite energies are set to $0$, and open boundary conditions are used.}
	\label{fig1}
\end{figure}

FBs typically emerge from destructive interference in certain lattice geometries~\cite{BAV,BAV2,BAV3}, such as Lieb~\cite{Lieb1,Lieb2}, diamond~\cite{diamond1,diamond2}, stub~\cite{stub2,stub1}, Kagome~\cite{Kagome1,Kagome2}, or sawtooth~\cite{Sawtooth1,Sawtooth2} lattices.  This destructive interference leads to the formation of compact localized states (CLSs). Given a single CLS, the whole CLS set can be generated through lattice translations. The CLS can be classified based on the minimum number $U$ of unit cells they span~\cite{CLS1,CLS2}. As shown in Fig.~\ref{fig1}, the cross-stitch lattice and sawtooth lattice are examples corresponding to $U=1$ and $U=2$, respectively. In most systems, FBs coexist with dispersive (non-flat) bands, whose presence may significantly influence both experimental measurements and practical applications of FB properties. Therefore, suppressing the influence of dispersive bands is both fundamentally and experimentally important.

Dissipation occurs widely in various systems and profoundly impacts the properties of these systems. Although dissipation is generally considered harmful to quantum correlations, recent progress in experimental techniques for controlling various types of dissipation~\cite{technology0,technology1,technology2,technology3,technology4,technology5,technology6,technology7,technology8,technology13,technology14,technology15,technology20,technology21} has led to growing interest in using dissipation to control quantum states or phase transitions~\cite{technology3,technology4,technology5,technology6,technology7,technology8,technology13,technology14,technology15,technology20,technology21,phenoma1,phenoma2,phenoma3,phenoma4,phenoma4s,phenoma5,phenoma6,PT0,PT1,PT2,Loc2,PT5,PT6,PT10,PT12}. 
In this Letter, we investigate a type of experimentally realizable dissipation applied to systems with dispersive and FBs, and show that it can drive the system's steady-state occupation into the FB, regardless of the initial state. This indicates that such dissipation can transfer particles from the dispersive bands into the FB, thereby serving as a control mechanism to eliminate the influence of dispersive bands. This control effect is applicable to a variety of FB systems.


{\em General discussion on the conditions for bond dissipation driving particles into the flatband.---} 
The dynamical evolution of an open quantum system under Markov dissipation  
follows the Lindblad master equation~\cite{GLindblad,HPBreuer},
\begin{align}
	\frac{d\rho}{dt}=\mathcal{L}[\rho] =-i[H,\rho]+\Gamma\sum_{j}( O_{j}\rho O_{j}^{\dagger}-\frac{1}{2}\{
	O_{j}^{\dagger}O_{j},\rho\}) , \notag 
\end{align}
where $\mathcal{L}$ represents the Lindbladian, with the steady-state density matrix $\rho_{ss}$
corresponding to its eigenstate with a zero eigenvalue, i.e., $\mathcal{L}[\rho_{ss}] = 0$.
$O_j$ represents the jump operator that acts on a pair of sites $j$ and $j+q$, with a site-independent dissipation strength $\Gamma$, as defined by~\cite{phenoma1,PT10,PZollerB1,PZollerB2,PZollerB3,PZollerB4,Marcos2012,Yusipov1,Yusipov2,YuchengW1,YuchengW2}:
\begin{equation}
	O_{j}=(c_{j}^{\dag}+ac_{j+q}^{\dag})(c_{j}-ac_{j+q}), \label{eq_oj}%
\end{equation}
where $a=\pm 1$ and $q\geq 1$. 
This dissipation, termed bond dissipation, preserves particle number but modifying the relative phase between pairs of sites separated by a distance $q$, favoring in-phase ($a=1$) or out-of-phase ($a=-1$) configurations. Such dissipation has been proposed for implementation in various experimental platforms, including cold atoms~\cite{phenoma1,PT10,PZollerB1,PZollerB2,PZollerB3,PZollerB4}, superconducting microwave resonator arrays~\cite{Marcos2012}, and superconducting quantum circuits~\cite{ChenS}.

We now discuss the conditions for constructing dark states of bond dissipation in FBs.
For an eigenstate $|\Psi_n\rangle$, if it satisfies the condition that a set of dissipative operators $O_j$ annihilate it, i.e.,
\begin{equation}
	\forall j,\quad  O_j|\Psi_n\rangle = 0, 
	\label{eq_dark}
\end{equation}
then $|\Psi_n\rangle$ is a dark state~\cite{PZollerB1}. If such a dark state $|\Psi_n\rangle$ exists, the steady state can be a pure state $\rho_{ss}=|\Psi_n\rangle\langle\Psi_n|$.
In a $U$-class FB system with $L$ unit cells, there are $N=L/U$ CLSs. Any eigenstate on the FB can be constructed through the superposition of these $N$ CLSs, i.e.,
\begin{equation}
	|\Psi_n\rangle = \sum_{j=1}^{N} A_j |\phi_{\text{CLS}}^{j}\rangle,
	\label{eq_FB}
\end{equation}
where $A_j$ denotes the complex amplitude for the $j$-th CLS $|\phi_{\text{CLS}}^{j}\rangle$.
We first treat a CLS as a whole, namely, $c^{\dagger}_j$ ($c_j$) in Eq. (\ref{eq_oj}) represents the creation (annihilation) operator of the $j$-th CLS. From Eqs. (\ref{eq_oj}, \ref{eq_dark}, \ref{eq_FB}), it is straightforward to see that when $A_j$, $|\phi_{\text{CLS}}^{j}\rangle$, and $q$ satisfy the following three conditions:
\begin{equation}
	A_{j} = A_{j+\kappa}, \quad |\phi_{\text{CLS}}^{j+\kappa}\rangle=a|\phi_{\text{CLS}}^{j}\rangle, \quad\text{and} \quad  q = \kappa,  
	\label{condition}
\end{equation}
where $\kappa\ge 1$ is an integer, the FB state in Eq.(\ref{eq_FB}) will become a dark state that satisfies Eq.(\ref{eq_dark}). 
In FB systems, lattice translations of a CLS produce spatially disjoint and phase-independent copies, allowing their relative phases to be freely chosen. This freedom enables matching the phase structure required by bond dissipation, such as assigning $\pm 1$ between CLSs. Since any linear combination of these CLSs remains an eigenstate of the flat band, the coefficients $A_j$ can be tuned accordingly. 
Thus, the first two conditions in Eq.(\ref{condition}) can be satisfied for any sign of $a$. 
In multi-chain systems, satisfying the third condition above, $q=\kappa$, means that $q$ spans $\kappa U$ unit cells on each chain, and that the bond dissipation must take the same form on all chains.
That is, the conditions for CLSs on a FB to form a dark state under bond dissipation are:
\begin{equation}
  q = \kappa U, \quad  \text{and} \quad O^{1}_j=O^{\alpha}_j (\alpha > 1),
	\label{condition2}
\end{equation}
where $O^{\alpha}_j$ denotes the bond dissipation operator on the $\alpha$-th chain, which must match that on the first chain, $O^{1}_j$. This constraint arises because the amplitude and phase relationships among lattice sites within each CLS are fixed. The bond dissipation on one chain determines the internal phase relationships, which in turn imposes consistent phase conditions on the other chains to preserve the CLS pattern. Thus, the form of dissipation must be identical across chains. When Eq.~(\ref{condition2}) is satisfied, the system evolves to a steady state occupying only the FB region. In certain cases, however, these requirements can be relaxed, as discussed below: 1. Eq.(\ref{condition2}) does not require the dissipation strength $\Gamma$ to be the same across all chains.
In many cases, applying bond dissipation on only one chain ( 
$\Gamma=0$ on the others) is sufficient to reach the FB steady state (see Supplemental Material \cite{SM}).  2. The requirement on 
$q$ in Eq. (\ref{condition2}) can be relaxed for some specific CLS distributions \cite{SM}. 

When bond dissipation satisfies the conditions in Eq. (\ref{condition2}), the system evolves into a steady state fully occupying the FB, regardless of the initial state. We illustrate this using the cross-stitch and sawtooth lattices [Fig.~\ref{fig1}]. Both models consist of upper and lower chains, with each unit cell containing two lattice sites. The wave function is written as $|\Psi\rangle=(\dots,\psi_{j-1},\psi_j,\dots)$, where each $\psi_j$ is a two-component vector representing the amplitudes on the upper and lower sites of the $j$-th unit cell. Using the eigenvalue equation 
$H\Psi=E\Psi$, we obtain:
\begin{equation}
	H_{-1}\psi_{j-1}+H_0\psi_j+H_1\psi_{j+1}=E\psi_j,
	\label{ham}
\end{equation}
where the $2\times 2$ matrices $H_1=H_{-1}^{\dagger}$ describe the hopping between the lattice sites of neighboring unit cells, and $H_0$ describes the on-site potentials and the hopping between lattice sites within the same unit cell.

{\em Cross-stitch model.---} We first apply bond dissipation to the cross-stitch lattice, a $U=1$ FB system with one flat and one dispersive band [Figs.~\ref{fig1}(a) and (b)]. The matrices $H_0$ and $H_1$ in Eq.(\ref{ham}) are given by
\begin{equation}
	H_0 = -t_0
	\begin{pmatrix}
		0 & 1 \\
		1 & 0
	\end{pmatrix},\quad\quad 
	H_1 = -t_1
	\begin{pmatrix}
		1 & 1 \\ 
		1 & 1
	\end{pmatrix},
	\label{eq_H_cs}
\end{equation}
where $t_0$ and $t_1$ denote the intra-cell and inter-cell hopping amplitudes, respectively.
Without loss of generality, we set $t_1=1$ and $t_0=0$ [Fig.~\ref{fig1}(a)], such that the FB lies at energy $E_{FB}=0$ [Fig.~\ref{fig1}(b)]. This FB hosts CLSs of the form $|\phi_{\text{CLS}}\rangle= (1, -1)^T$ [Fig.~\ref{fig1}(a)], strictly localized within a single unit cell.

\begin{figure}[t]
	\centerline{\includegraphics[width=1\linewidth]{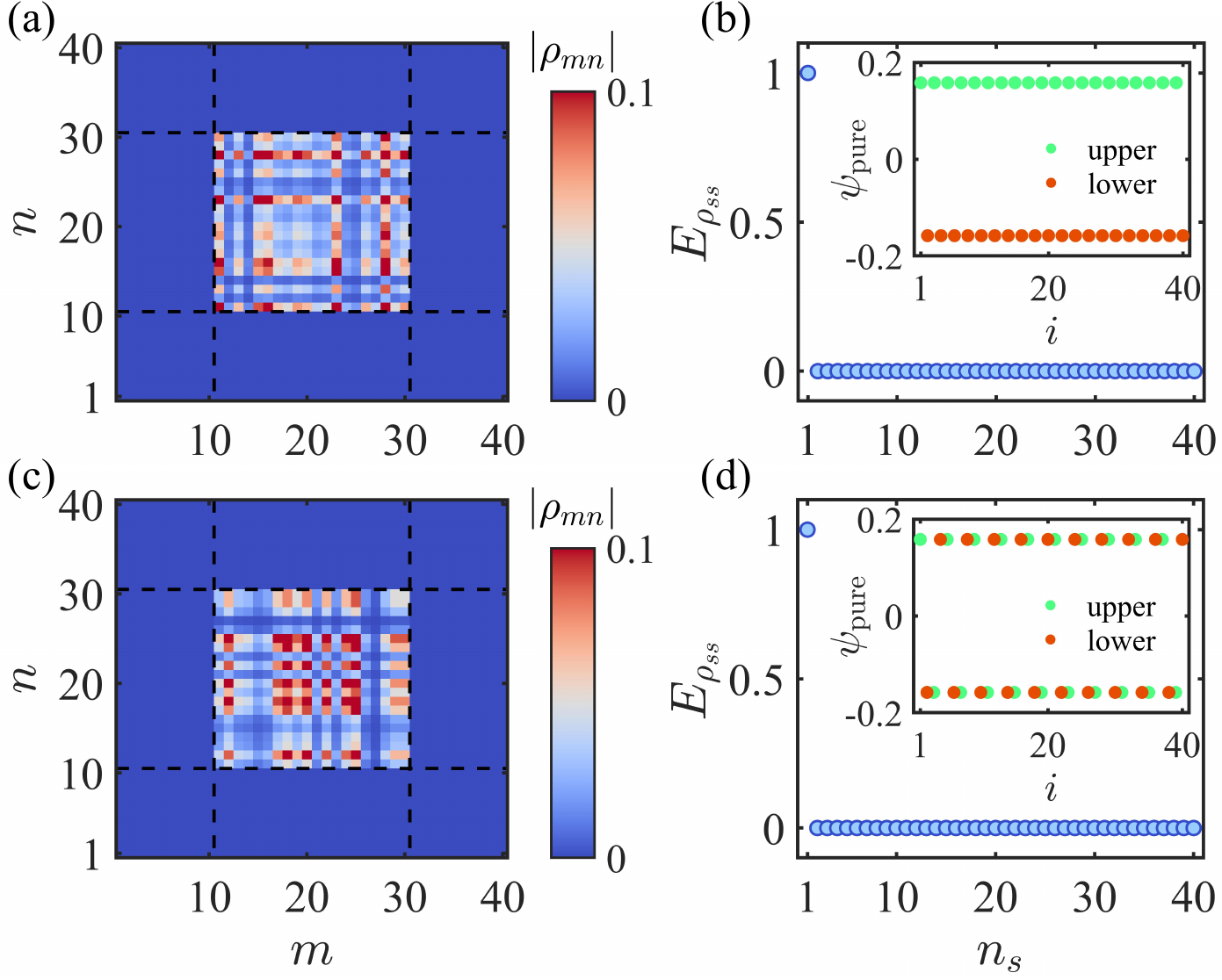}}
	\caption{Steady-state density matrix $\rho_{ss}$ of the cross-stitch model under bond dissipation with 
		$q=1$ and $O_{j}^{u}=O_{j}^{l}$. (a, c): Matrix elements $|\rho_{mn}|$ in the eigenbasis of $H$ for $a=1$ and $a=-1$, respectively. Dashed lines separate the FB region in the center from the dispersive bands on the sides. (b, d): Eigenvalue spectra of $\rho_{ss}$ for $a=1$ and $a=-1$, respectively, each showing a single nonzero value, confirming a pure steady state. Insets: real-space wavefunctions corresponding to the nonzero eigenvalue. Here, we consider $20$ unit cells and set $\Gamma=1$ on both chains. }
	\label{fig2}
\end{figure}

As specified in Eq. (\ref{condition2}), the jump operators used here satisfy two conditions:
(1) For the cross-stitch lattice, which belongs to the $U=1$ class model, the requirement reduces to $q\geq 1$; (2) The bond dissipation operators are identical on both chains, i.e., $O_{j}^{u}=O_{j}^{l}$, where $O_{j}^{u}$ (or $O_{j}^{l}$) represents the jump operator acting on the upper (or lower) chain. The effects of $O_{j}^{u}\neq O_{j}^{l}$ are discussed in the Supplemental Material~\cite{SM}. Our results are independent of the dissipation strength $\Gamma$, which can vary across chains and even vanish on one chain. Without loss of generality, we set $\Gamma=1$ for both chains.  We first fix $q=1$ and $a=1$, and examine the steady state $\rho_{ss}$ in the eigenbasis of the Hamiltonian 
$H$, with matrix elements $\rho_{mn}=\langle\Psi_m|\rho_{ss}|\Psi_n\rangle$, where $|\Psi_m\rangle$
and $|\Psi_n\rangle$ represent the eigenstates of $H$. Fig.~\ref{fig2}(a) shows that $\rho_{ss}$ is entirely confined to the FB sector, independent of the initial state, confirming that the bond dissipation drives population from the dispersive band into the FB. Diagonalizing $\rho_{ss}$ yields a single nonzero eigenvalue [Fig.~\ref{fig2}(b)], confirming that the steady state is pure. The inset in Fig.~\ref{fig2}(b) displays the real-space wave function distribution of this pure state, where $i=2j-1$ (or $2j$) corresponds to the upper (or lower) chain of the $j$-th unit cell, showing the out-of-phase on the upper and lower chains within the same unit cell (consistent with the distribution of the CLS in the FB, $|\phi_{\text{CLS}}\rangle = (1, -1)^T$) while maintaining the in-phase across different unit cells on the same chain, a behavior resulting from setting $a=1$ in the bond dissipation. Therefore, this pure state is constructed from an equal-weight superposition of CLSs of the cross-stitch model: $|\Psi_{\text{pure}}\rangle = \sum_{j}^{L} A |\phi^{j}_{\text{CLS}}\rangle$.
When $a=-1$, the steady state of the system remains in the FB region [Fig.~\ref{fig2}(c)], and $\rho_{ss}$ still corresponds to a pure state [Fig.~\ref{fig2}(d)]. However, this pure state exhibits alternating phases between neighboring unit cells on the same chain, as shown in the inset of Fig.~\ref{fig2}(d), which arises because the bond dissipation with 
$a=-1$ selects the out-of-phase. In this case, the pure state can be expressed in terms of CLSs as:
$|\Psi_{\text{pure}}\rangle = \sum_{j}^{L} (-1)^{j} A |\phi^{j}_{\text{CLS}}\rangle$.

{\em Sawtooth model.---} We next take the sawtooth model as an example to study the effect of bond dissipation on $U=2$-type FB systems. When the onsite energies are the same (set to $0$ here), and the ratio of the diagonal hopping between the upper and lower chains to the baseline hopping within the lower chain is $\sqrt{2}$ [see Fig.~\ref{fig1}(c)], i.e., the system's Hamiltonian is given by~\cite{CLS1,CLS2}
\begin{equation}
	H_0 = -
	\begin{pmatrix}
		0 & \sqrt{2} \\
		\sqrt{2} & 0
	\end{pmatrix},\quad\quad 
	H_1 = -
	\begin{pmatrix}
		0 & \sqrt{2} \\
		0 & 1
	\end{pmatrix},
	\label{eq_H_st}
\end{equation}
this configuration results in a FB, as shown in Fig.~\ref{fig1}(d).
The FB contains a series of CLSs, each of which occupies two unit cells, given by
$|\phi_{\text{CLS}}\rangle = ((1, 0)^T, (1, -\sqrt{2})^T)$ [Fig.~\ref{fig1}(c)]. 

\begin{figure}[t]
	\centerline{\includegraphics[width=0.98\linewidth]{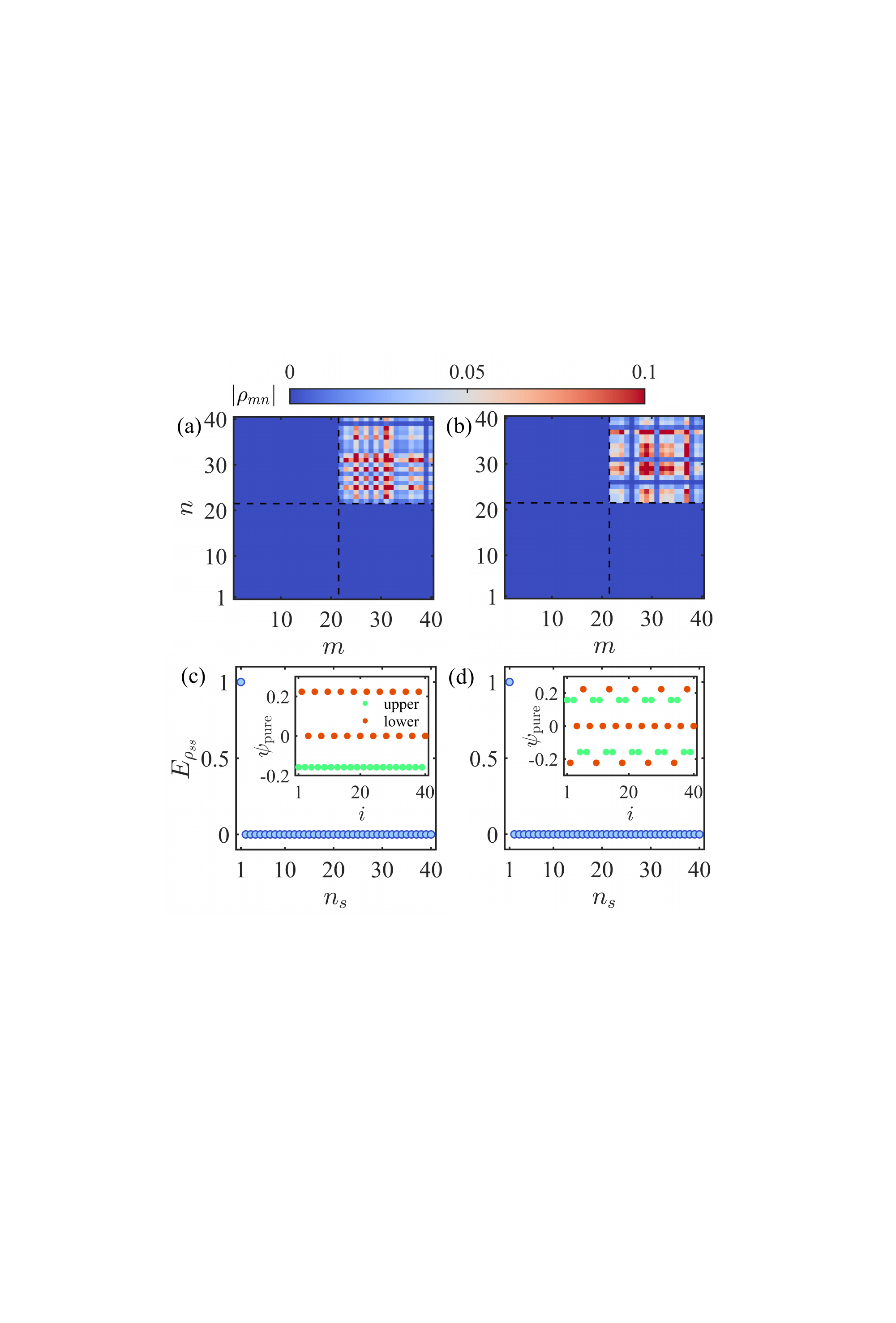}}
	\caption{Steady-state density matrix $\rho_{ss}$ of the sawtooth model under bond dissipation with 
		$q=2$ and $O_{j}^{u}=O_{j}^{l}$.  Matrix elements $|\rho_{mn}|$ in the eigenbasis of $H$ for (a) $a=1$ and (b) $a=-1$, respectively. Dashed lines mark the boundary between the FB and dispersive bands. Eigenvalue spectra of $\rho_{ss}$ for (c) $a=1$ and (d) $a=-1$, both featuring a single nonzero eigenvalue. The real-space distribution of the corresponding eigenstate is shown in the insets. Here, we fix $L=20$.}
	\label{fig3}
\end{figure}

Since the sawtooth model belongs to the $U=2$ class, according to Eq. (\ref{condition2}), the form of the applied bond dissipation satisfies $q=2\kappa$. Additionally, another requirement is the same as the second condition for the bond dissipation applied to the cross-stitch model: $O_{j}^{u}=O_{j}^{l}$. We first fix $q=2$ (i.e., $\kappa=1$). Figures \ref{fig3}(a) and \ref{fig3}(b) display the steady-state density matrix in the Hamiltonian's eigenbasis for 
$a=1$ and $a=-1$, respectively, demonstrating that the system's steady state always occupies the FB region. Moreover, in both cases, the steady-state density matrix $\rho_{ss}$
exhibits only one nonzero eigenvalue [see Figs. \ref{fig3}(c) and \ref{fig3}(d)], indicating that the final steady state is a pure state. For $a=1$, the pure state exhibits an in-phase distribution between next-nearest-neighbor lattice sites on the same chain (see inset in Fig. \ref{fig3}(c)), while for $a=-1$, it shows out-of-phase distribution (see inset in Fig. \ref{fig3}(d)).
Combining with the form of the CLS in this model, these two pure states can be respectively expressed in terms of CLS as:$|\Psi_{\text{pure}}\rangle = \sum_{j}^{N} A |\phi^{j}_{\text{CLS}}\rangle$ and $|\Psi_{\text{pure}}\rangle = \sum_{j}^{N}(-1)^j A |\phi^{j}_{\text{CLS}}\rangle$, where $N=L/U$ represents the total number of CLS.

\begin{figure}[t]
	\centerline{\includegraphics[width=1\linewidth]{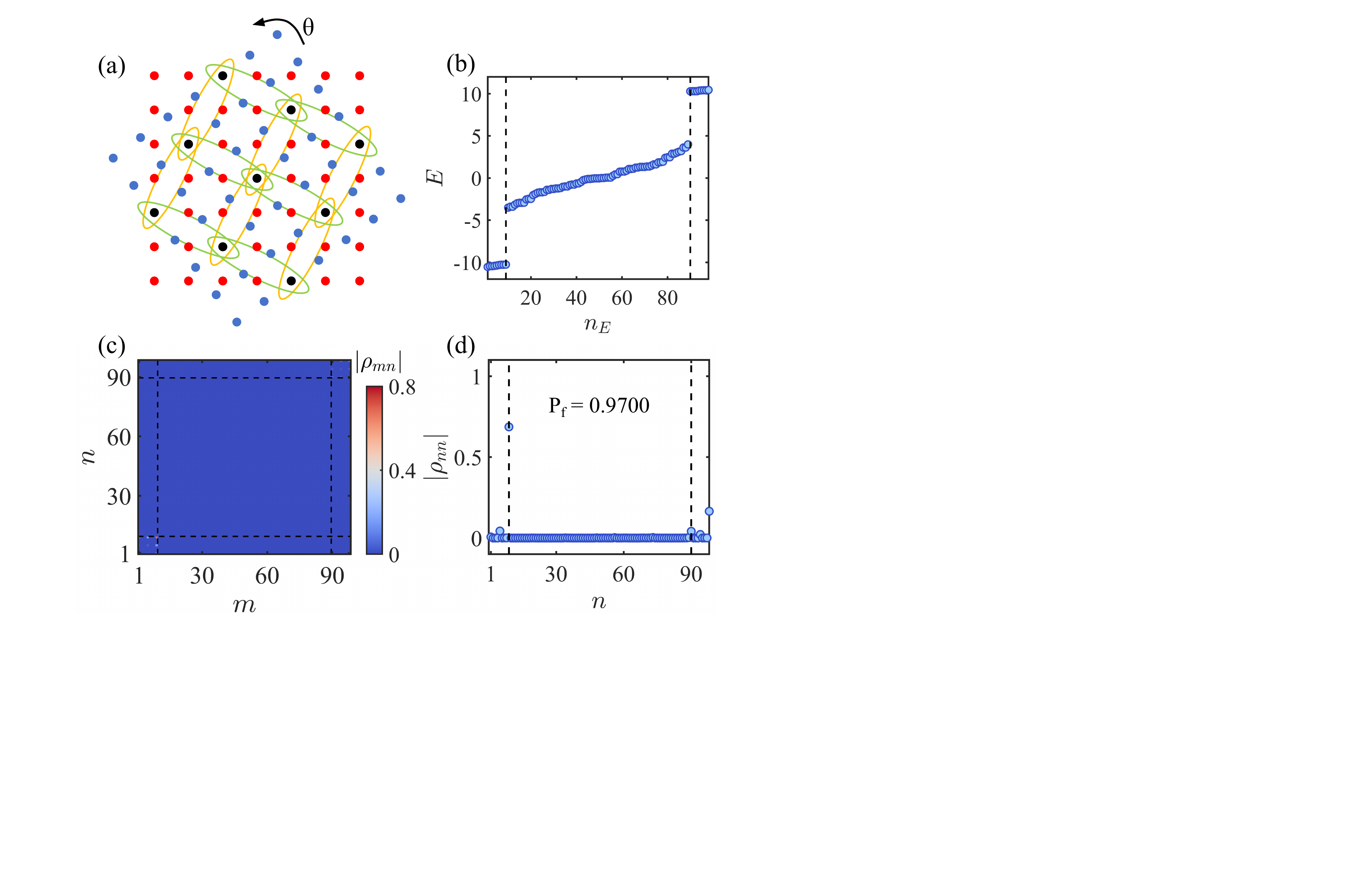}}
	\caption{(a) Twisted-bilayer square lattice with rotation angle $\theta$. Red (blue) dots represent sites in layer $1$ ($2$), and black dots indicate superlattice sites where the two layers align vertically. (b) Single-particle spectrum at $\theta = 36.87^\circ$ for $t_{\perp} = 10$, $\Delta = 10$, and $l_0 = 0.15$.
	 Bond dissipation is applied between neighboring black sites on layer $1$ only. The results are the same along the directions indicated by green and yellow lines. (c) Steady-state density matrix $|\rho_{mn}|$ in the eigenbasis of $H$ for $a=-1$. The diagonal elements are shown in (d). The dashed lines separate the nearly FB region from the dispersive band region. Here, we consider each layer to consist of a $7 \times 7$ lattice.}
	\label{fig4}
\end{figure}

{\em Twisted-bilayer square lattice.---} We finally examine a two-dimensional (2D) example: a twisted-bilayer square lattice, formed by rotating one square lattice relative to the other by an angle $\theta$ [see Fig.\ref{fig4}(a)]. The system is described by the Hamiltonian~\cite{flatPRA},
\begin{equation}
\begin{aligned}
	H ={} & -\Delta \sum_{\alpha,j} (-1)^{\alpha} \, c_{\alpha,j}^\dagger c_{\alpha,j}
	- t \sum_{\alpha=1,2} \sum_{\langle j,j'\rangle} c_{\alpha,j}^\dagger c_{\alpha,j'} \\
	& - \sum_{j,j'} t_{\perp}(j,j') \left( c_{1,j}^\dagger c_{2,j'} + \text{H.c.} \right),
\end{aligned}
\end{equation}
where $c_{\alpha,j}^\dagger$ ($c_{\alpha,j}$) creates (annihilates) a particle at site $j$ in layer $\alpha = 1, 2$, $\Delta$ is a staggered potential, $t$ is the intralayer hopping amplitude, and the interlayer tunneling takes the form $t_\perp(j,j') = t_{\perp} e^{-S_{jj'}^2/4l_0^2}$, with $S_{jj'}$ denoting the distance between sites $j$ and $j'$, and $l_0$ a Gaussian width parameter. This system forms a moir\'{e} superlattice with nearly FBs [Fig.\ref{fig4}(b)] and can be realized in optical lattice experiments~\cite{flatPRA,flatPRA2,flatnature} with all parameters being experimentally tunable. In the following calculations, we set $t = 1$, $t_{\perp} = 10$, $\Delta = 10$, $l_0 = 0.15$ and $\theta=36.87^\circ$~\cite{flatPRA}.

For simplicity and experimental feasibility, we apply bond dissipation only on layer $1$, between neighboring superlattice sites (indicated by black dots in Fig.\ref{fig4}(a)). These black dots represent vertically aligned sites across the two layers and define the moir\'{e} superlattice. We consider bond dissipation with parameters \( q = 1 \) and \( a = -1 \), favoring out-of-phase configurations. Bond dissipation can be applied along either the green or yellow directions shown in Fig.\ref{fig4}(a), both leading to similar results. As shown in Fig.\ref{fig4}(c) and \ref{fig4}(d), the steady-state density matrix is largely confined to the nearly FB sector. The FB occupation probability, $P_f$, defined as the sum of  diagonal elements over FB eigenstates, reaches $0.970$--despite the FB subspace occupying only a small fraction of the total Hilbert space. Choosing $a=1$ yields similar results as well~\cite{SM}. These results confirm that the mechanism of using dissipation to drive particles into FBs remains effective in complex 2D moir\'{e} systems.

\begin{figure}[t]
	\centerline{\includegraphics[width=1\linewidth]{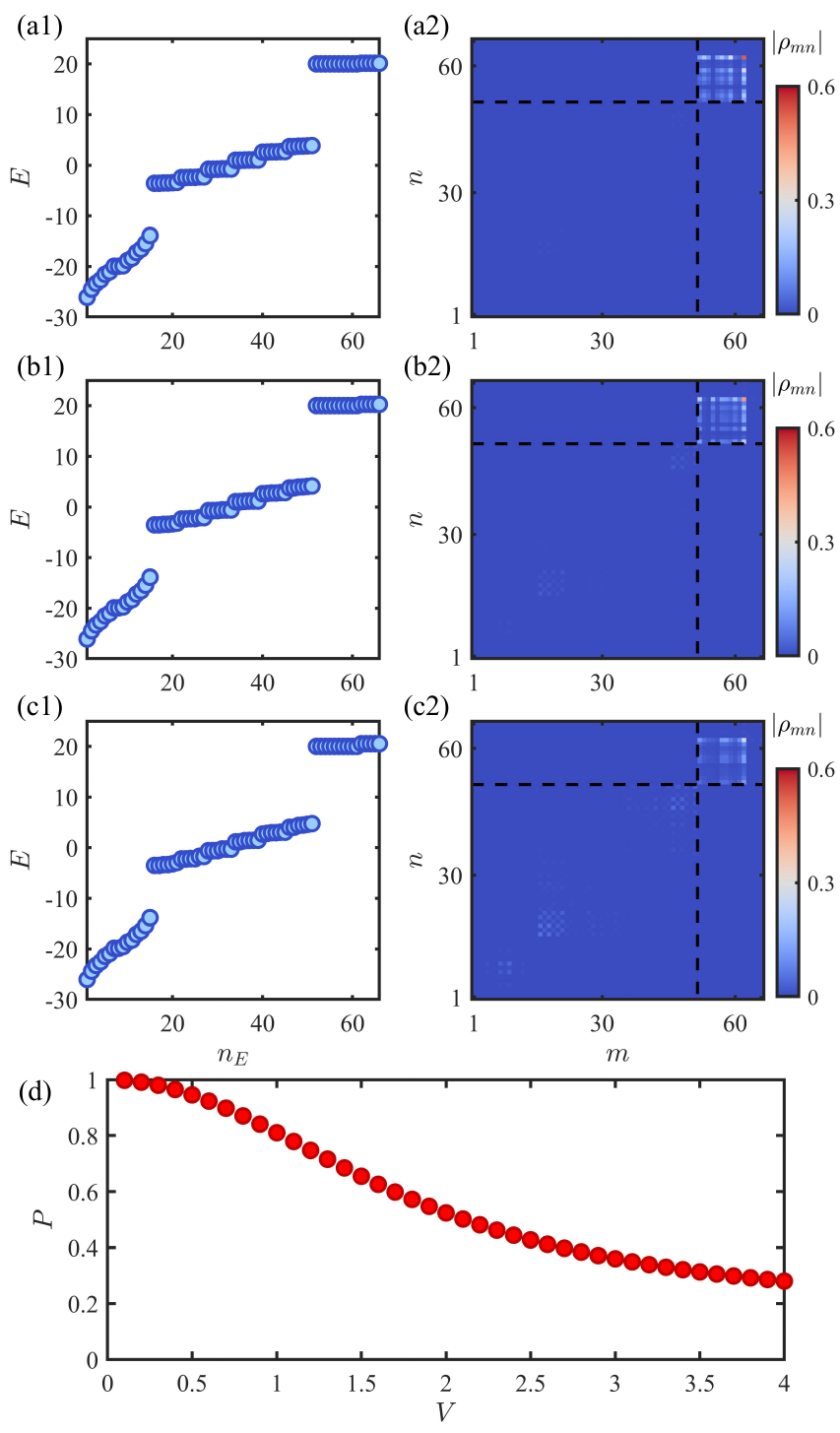}}
	\caption{(a1, b1, c1) Many-body energy spectra for interaction strengths 
			$V=0.5$, $1$, and $2$, respectively. (a2, b2, c2) Corresponding steady-state density matrix elements $|\rho_{mn}|$ in the eigenbasis of the many-body Hamiltonian.
			(d) Occupation $P$ as a function of interaction strength $V$. Here, we fix the number of unit cells to 
			$N=6$, the particle number to $2$, and set $t_0=10$, $t_1=1$.}
	\label{fig6}
\end{figure}

{\em Effects of interactions on dissipation-driven flat band occupation.---} We investigate how interparticle interactions affect the steady-state occupation of FBs induced by bond dissipation, using the cross-stitch lattice model. The system Hamiltonian is given by $H=H_0+H_1+H_{int}$, where $H_0$ and $H_1$ describe intra-cell and inter-cell hopping, respectively, with matrix forms given in Eq.~(7). The interaction term is $H_{int}=V\sum_jn_jn_{j+1}$, representing nearest-neighbor interactions with strength $V$.
In the following, we fix $t_1=1$, $t_0=10$ (so that the single-particle flat band lies at $E_{\text{FB}}=10$), and apply identical bond dissipation on the upper and lower chains with parameters 
$q=2$, $a=1$, and $\Gamma=1$. We then study how increasing $V$ influences the steady state. Hereafter, we fix the number of unit cells to 
$N=6$ (i.e., system size $12$) and consider two particles. As shown in Fig.~\ref{fig6} (a1), the FB structure near the top of the spectrum remains well preserved under weak interaction. When bond dissipation is applied, the steady state is almost fully confined to the FB region [Fig.~\ref{fig6} (a2)]. As the interaction strength increases, the FB becomes slightly distorted [Fig.~\ref{fig6} (b1)], yet the steady state remains primarily localized within the nearly FB region [Fig.~\ref{fig6} (b2)].  
With stronger interactions, the FB characteristics deteriorate further [Fig.~\ref{fig6} (c1)], leading to reduced occupation in the original FB region [Fig.~\ref{fig6} (c2)]. To describe this transition process, we introduce the quantity $P=\sum_{i=D_H-N_p+1}^{D_H} |\rho_{ii}|$, where $N_p$ is the number of FB states located at the top of the spectrum in the noninteracting case, and 
$\rho_{ii}$ denotes the diagonal elements of the steady-state density matrix, sorted in ascending order according to the eigenvalues $E_i$ of the Hamiltonian. Thus, $P$ represents the total population in the upper-right region of the diagonal, separated by dashed lines in Figs.~\ref{fig6}(a2, b2, c2). It is worth noting that as the interaction strength increases, this region may no longer correspond to a true FB. Therefore, the definition of 
$P$ here differs slightly from $P_f$ used in Fig.~4. As shown in Fig.~\ref{fig6}(d), when the interaction is weak, 
$P$ is close to $1$, indicating nearly complete occupation of the FB sector. As the interaction becomes stronger, the FB character degrades and $P$ gradually decreases. In FB systems, where the kinetic energy is nearly quenched, even weak interactions can lead to strong correlation effects. Consequently, bond dissipation can also drive particles from dispersive bands into FBs in strongly correlated systems, as long as the interactions do not significantly disrupt the FB structure.

{\em Conclusion and Discussion.---} We have shown that bond dissipation satisfying Eq. (\ref{condition2}) can drive particles from dispersive bands into FBs. Once the steady state is reached, dissipation can be turned off, and the system evolves unitarily as $\rho(t)=\sum_{nm}e^{i(E_n-E_m)t}\rho_{nm}|\Psi_n\rangle\langle\Psi_m|$, where 
 $|\Psi_n\rangle$ and $|\Psi_m\rangle$ are eigenstates of the Hamiltonian 
$H$, with eigenvalues $E_n$ and $E_m$, respectively. The diagonal elements $\rho_{nn}(t)=\rho_{nn}$ remain unchanged, indicating that the system remains confined to the FB sector even after dissipation is removed. 
Thus, dissipation acts as a tool to transfer particles from dispersive bands into FBs, without modifying the Hamiltonian itself.

Our mechanism can be readily tested using cold atoms in optical lattices, where numerous FB models have already been realized~\cite{cold1,cold2,cold3,cold4,cold5,cold6,cold7,cold8}, including the cross-stitch ~\cite{cold3}, sawtooth~\cite{cold2} and twisted-bilayer square lattices~\cite{flatnature} discussed in this work. Moreover, the implementation of bond dissipation [Eq.(\ref{eq_oj})] was originally proposed in cold-atom setups~\cite{phenoma1,PZollerB1,PZollerB2,PZollerB3,PZollerB4}. The initial approach involved introducing a driving laser to achieve the $q=1$ case by adjusting the relationship between the wavelength of the driving laser and the optical lattice. More recent schemes using polarization modulation allow for realizing larger $q$ values, including $q=2$ and beyond~\cite{PT10}. 

By eliminating the influence of dispersive-band states, dissipation enables controlled preparation and exploration of FB physics, with potential applications in strongly correlated systems across theory, experiment, and materials design. Our work provides a new route to utilize dissipation for driving particles into FBs, offering fresh insights into the study of FBs in open systems. 

\begin{acknowledgments}
This work is supported by National Key R\&D Program of China under Grant No.2022YFA1405800, the Key-Area Research and Development Program of Guangdong Province (Grant No.2018B030326001), Guangdong Provincial Key Laboratory(Grant No.2019B121203002). C. Y. acknowledges support from the Key-Area Research and Development Program of Guangdong Province Grant No.2020B0303010001, Grant No.2019ZT08X324, No.2019CX01X042, No.2019B121203002.
\end{acknowledgments}
	

\global\long\def\id{\mathbbm{1}}
\global\long\def\ui{\mathbbm{i}}
\global\long\def\ud{\mathrm{d}}


\global\long\def\id{\mathbbm{1}}
\global\long\def\ui{\mathbbm{i}}
\global\long\def\ud{\mathrm{d}}
\setcounter{equation}{0} \setcounter{figure}{0}
\setcounter{table}{0} 
\renewcommand{\theparagraph}{\bf}
\renewcommand{\thefigure}{S\arabic{figure}}
\renewcommand{\theequation}{S\arabic{equation}}

\onecolumngrid
\flushbottom
\newpage
\section*{\large Supplementary Material:\\Dissipation-Driven Transition of Particles from Dispersive to Flat Bands}
In the main text, we require identical forms of bond dissipation for both the upper and lower chains, i.e., $Q_j^{u}=Q_j^{l}$, while allowing the dissipation strengths $\Gamma$ of the two chains to differ.
In the Supplementary Materials, we first examine the physical consequences arising from distinct bond dissipation configurations between the upper and lower chains. We then investigate the physical effects when bond dissipation is applied to only a single chain, and further examine the emergence of multiple steady states when $q>1$ in the bond dissipation. Finally, we investigate the effect of bond dissipation applied to a two-dimensional (2D) flat-band (FB) system.
In Sections (I) and (II), we fix the dissipation strength $\Gamma$ on both chains to be $1$. We denote the elements in the bond dissipation operator $O^u$ applied to the upper chain as $a^u$ and $q^u$, and those in the bond dissipation operator $O^l$ applied to the lower chain as $a^l$ and $q^l$.
In Section (III), we consider a one-dimensional (1D) Lieb lattice, which consists of three coupled chains. We define the bond dissipation operator on the middle chain as $O^m$, with parameters $a^m$ and $q^m$.

\subsection*{I. $a^{u}\neq a^{l}$}
\begin{figure}[h]
	\centerline{\includegraphics[width=0.4\linewidth]{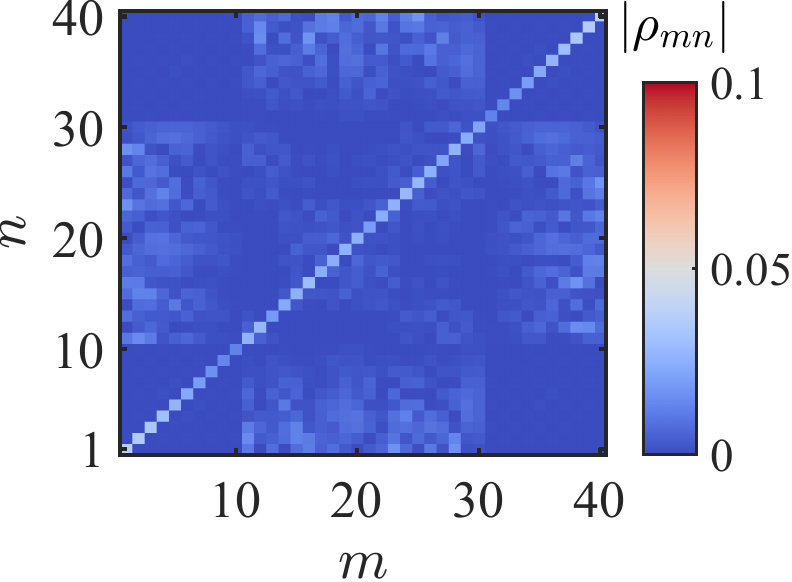}}
	\caption{The absolute values of the steady state's density matrix elements in the eigenbasis of $H$. We employ identical conditions and parameters as in Fig. 2 of the main text, except for modifying the previously equal chain coefficients to asymmetric values $a^{u}=1$ and $a^{l}=-1$.}%
	\label{figB1}%
\end{figure}

We take the cross-stitch model as an example, keeping $q$ of both the upper and lower chains fixed at $1$. However, the bond dissipation added to the upper chain is set to $a=1$, while the bond dissipation added to the lower chain is set to 
$a=-1$. Fig.\ref{figB1} demonstrates that the system's steady state does not exclusively occupy the FB region. This can be understood as follows: While different compact localized states (CLSs) may acquire arbitrary relative phases, the intra-CLS phase difference between upper and lower chains is fixed. For instance, the CLS in the cross-stitch model takes the form $|\phi_{\text{CLS}}\rangle= (1, -1)^T$, indicating opposite phases on the two chains. The condition $a^{u}=1$ requires identical phases between neighboring sites on the upper chain for steady-state formation. If the steady state were composed of CLS superpositions, this would further impose identical phases between lower-chain neighbors (as all lower-chain occupations must maintain $\pi$-phase opposition to the upper chain). However, $a^{l}=-1$ favors anti-phase configurations between neighboring sites, thereby disrupting the CLS construction. Consequently, the asymmetry condition $a^{u}\neq a^{l}$ prevents complete FB occupation in the final steady state.

\subsection*{II. $q^{u}\neq q^{l}$}
\begin{figure}[t]
	\centerline{\includegraphics[width=0.65\linewidth]{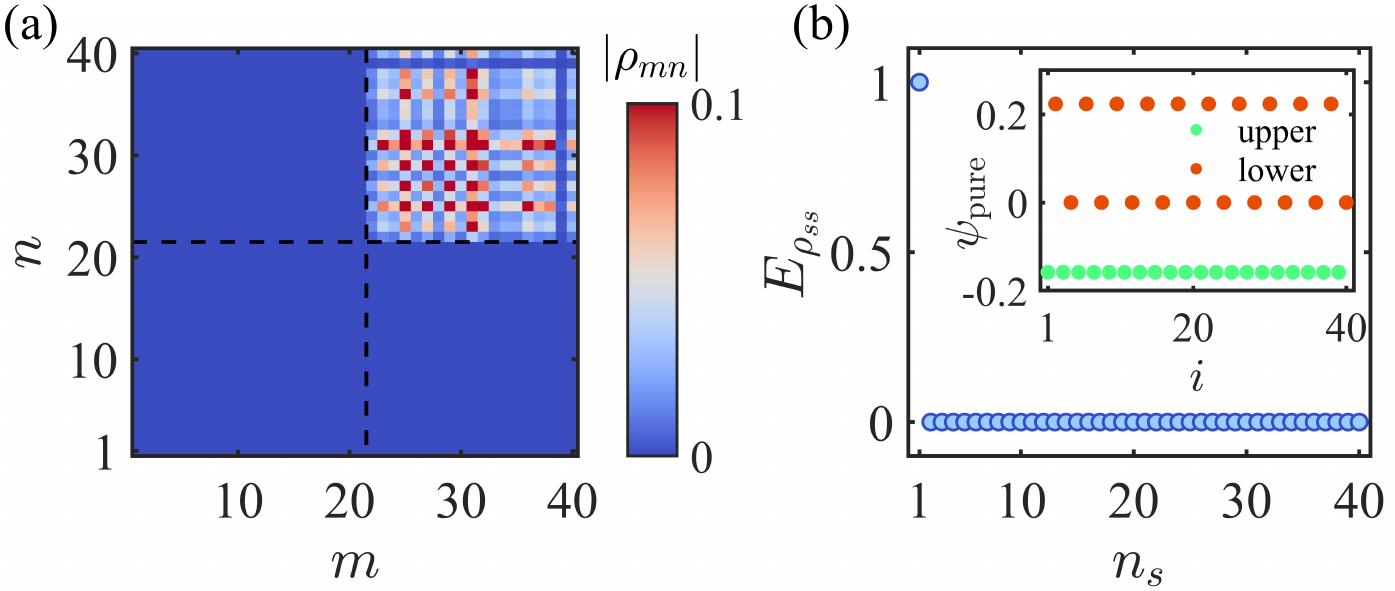}}
	\caption{ We adopt the same conditions and parameters as those in Fig. 4(a) of the main text, except that we modify the originally equal chain coefficients from $q^u=q^l=2$ to $q^u=1$ and $q^l=2$. Note that in this case, we fix 
		$a=1$. $a=-1$ cannot achieve a steady-state occupation in the FB.}%
	\label{figB2}%
\end{figure}

We take the sawtooth model as an example, keeping $a$ of both the upper and lower chains fixed at $1$. However, 
$q^u$ and $q^l$ are different, with $q^u=1$ and $q^l=2$. Fig. \ref{figB2}(a)(b) and Fig. 4(a)(c) in the main text are exactly the same. This system's steady state is occupied in the FB region and is a pure state, which also exhibits an in-phase distribution between next-nearest-neighbor lattice sites on the lower chain, while on the upper chain, neighboring lattice sites share the same phase distribution. Such a choice of bond dissipation depends on the configuration of the CLS in the sawtooth lattice. Each CLS occupies two unit cells, where the distribution within a CLS is such that neighboring lattice sites on the upper chain have the same phase (thus, when $q^u=1$, only $a=1$ is allowed, and $a=-1$ is not allowed), while on the lower chain, one site is occupied while the other remains unoccupied (see Fig.1(c) in the main text). 
Therefore, when the specific configuration of the CLS is known in advance, the form of bond dissipation can be designed based on the characteristics of the CLS, and it does not necessarily need to satisfy Eq. (5) in the main text. However, Eq. (5) in the main text is always capable of achieving a steady-state occupation in the FB, regardless of the model or the form of the CLS.

\subsection*{III. Bond dissipation is applied to only one chain}
\begin{figure}[t]
	\centering
	\includegraphics[width=\linewidth]{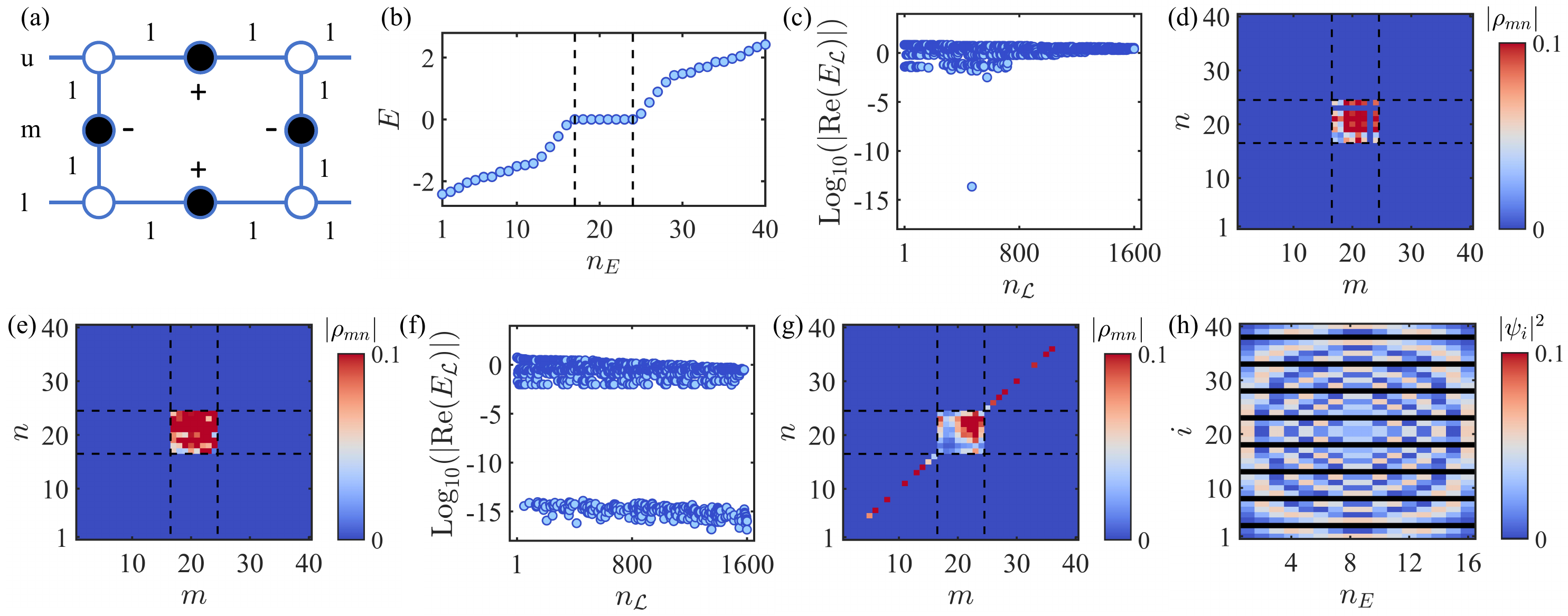}
	\caption{(a) 1D Lieb lattice with $U=2$ CLS configuration. Circles represent lattice sites, and filled circles indicate the spatial distribution of a CLS.  The hopping strength between lattice sites is set to 1. The corresponding energy spectrum is shown in (b), where a FB appears at $E=0$ in the middle of the spectrum. 
		(c) and (f): The absolute value of the real parts of the eigenvalues of the Liouvillian superoperator. (d), (e) and (g): The steady state density matrix elements in the eigenbasis of $H$, with dashed lines separating the FB region from the dispersive band region. We adopt the following forms of bond dissipation: (c) and (d): $q^u=q^l=4$, $q^m=2$, $\Gamma^u=\Gamma^l=\Gamma^m=1$, $a^u=a^l=a^m=1$; (e): $\Gamma^u=1$, $\Gamma^l=\Gamma^m=0$, $q^u=4$, $a^u=1$; (f) and (g): $\Gamma^m=1$, $\Gamma^u=\Gamma^l=0$, $q^m=2$, $a^m=1$. 
		(h) Real-space distribution of the 16 clearly nonzero states in the dispersive band region shown in (g). The horizontal axis represents the indices of these states, and the vertical axis $i$ denotes the real-space lattice sites.
		In the calculations for Figs. (b-h), we fix the number of unit cells to $L=8$ and use open boundary conditions.}
	\label{figS3}
\end{figure}

For many models that feature FBs, applying bond dissipation to only a single chain can still drive the system toward a steady FB state, as demonstrated in the two models discussed in the main text. However, in certain cases, due to the influence of specific eigenstates, bond dissipation on a single chain may not be sufficient to drive particles from the dispersive band into the FB. Here, we take the 1D Lieb lattice as an example. Its lattice structure is shown in Fig. \ref{figS3}(a), where each unit cell contains five sites. The hopping amplitude between sites is set to $1$. The Hamiltonian consists of intra-cell and inter-cell terms $H=H_{\text{intra}}+H_{\text{inter}}$. The intra-cell term is 
\begin{equation}
	H_{\text{intra}}=\sum_{n}(u^{\dagger}_{n,1}u_{n,2}+u^{\dagger}_{n,1}m_n+l^{\dagger}_{n,1}l_{n,2}+l^{\dagger}_{n,1}m_n+h.c.),
\end{equation}
and the inter-cell term is
\begin{equation}
	H_{\text{inter}}=\sum_{n}(u^{\dagger}_{n,2}u_{n+1,1}+l^{\dagger}_{n,2}l_{n+1,1}+h.c.).
\end{equation}
Here $u_{n,1}, u_{n,2}, m_n, l_{n,1}, l_{n,2}$ are the annihilation operators for the two upper chain sites ($u_{n,1}$, $u_{n,2}$), the middle chain site ($m_n$), and the two lower chain sites ($l_{n,1}$, $l_{n,2}$) in the 
$n$-th unit cell, respectively. It is easy to show that this model features a FB at 
$E=0$ [Fig. \ref{figS3}(b)], and the states on the FB form a $U = 2$ CLS configuration [Fig. \ref{figS3}(a)].

According to Eq. (5) in the main text (note: in Eq. (5), $q=\kappa U$, where the unit of $q$ is unit cells, and the index $j$ in $O_j^1=O_j^{\alpha}$  also refers to unit cells), applying bond dissipation at every site between next-nearest-neighbor unit cells can drive particles from the dispersive bands into the FB states.
Since each unit cell of the upper and lower chains contains two sites, we apply bond dissipation with $q^u=q^l=4$
to the upper and lower chains. The middle chain contains one site per unit cell, so the dissipation applied there is 
$q^m=2$. We fix the phase-tuning parameter as $a^u=a^l=a^m=1$, and set the dissipation strength to $\Gamma^u=\Gamma^l=\Gamma^m=1$. Under this dissipation scheme, the system has a unique steady-state [Fig. \ref{figS3}(c)], which exclusively selects the FB occupation states [Fig. \ref{figS3}(d)].

Next, we apply bond dissipation to only one chain. When bond dissipation is applied solely to the upper chain (with the dissipation strengths on the middle and lower chains set to $\Gamma^m=\Gamma^l=0$) or solely to the lower chain (
$\Gamma^m=\Gamma^u=0$), the system can still reach a steady state that exclusively occupies the FB region [Fig. \ref{figS3}(e)]. However, a different phenomenon emerges when bond dissipation is applied only to the middle chain (with the dissipation on the upper and lower chains turned off, i.e., $\Gamma^u=\Gamma^l=0$). In this case, the system exhibits a large number of steady states, as shown in Fig. \ref{figS3}(f). We focus on the steady state corresponding to the smallest $|\text{Re}(E_{\mathcal{L}})|$ shown in Fig. \ref{figS3}(f) and display the absolute values of its density matrix elements in the eigenbasis of the Hamiltonian in Fig. \ref{figS3}(g). It can be seen that this steady state partially occupies the FB and partially occupies the dispersive bands. To understand this further, we show the real-space distribution of the states that reside in the dispersive bands in Fig. \ref{figS3}(h). In our calculations, we fix the system size to $8$ unit cells, corresponding to $40$ lattice sites. Among the steady-state components shown in Fig. \ref{figS3}(g), there are $16$ states with clearly nonzero values located in the dispersive band region. As seen in Fig. \ref{figS3}(h), all $16$ of these states have zero occupation probability on the middle-layer sites. The black regions in the figure indicate near-zero occupation probabilities, and their positions can be expressed as $5(z-1)+3$, where $z$ is a positive integer (starting from $1$), $5$ corresponds to the number of sites per unit cell, and $3$ denotes the position of the middle-layer site within each unit cell. This occupation pattern is consistent with the phase selection of the bond dissipation. Thus, the bond dissipation applied to the middle chain selects these specific states as steady states. Therefore, when the applied bond dissipation takes the form given in Eq. (5) of the main text, the system's steady states are guaranteed to lie within the FB region. When bond dissipation is applied to only a single chain, in most cases it can still drive particles from the dispersive bands into FB, but some exceptions do exist.

\subsection*{IV. Multiple Steady States for $q>1$}

\begin{figure}[t]
	\centerline{\includegraphics[width=0.5\linewidth]{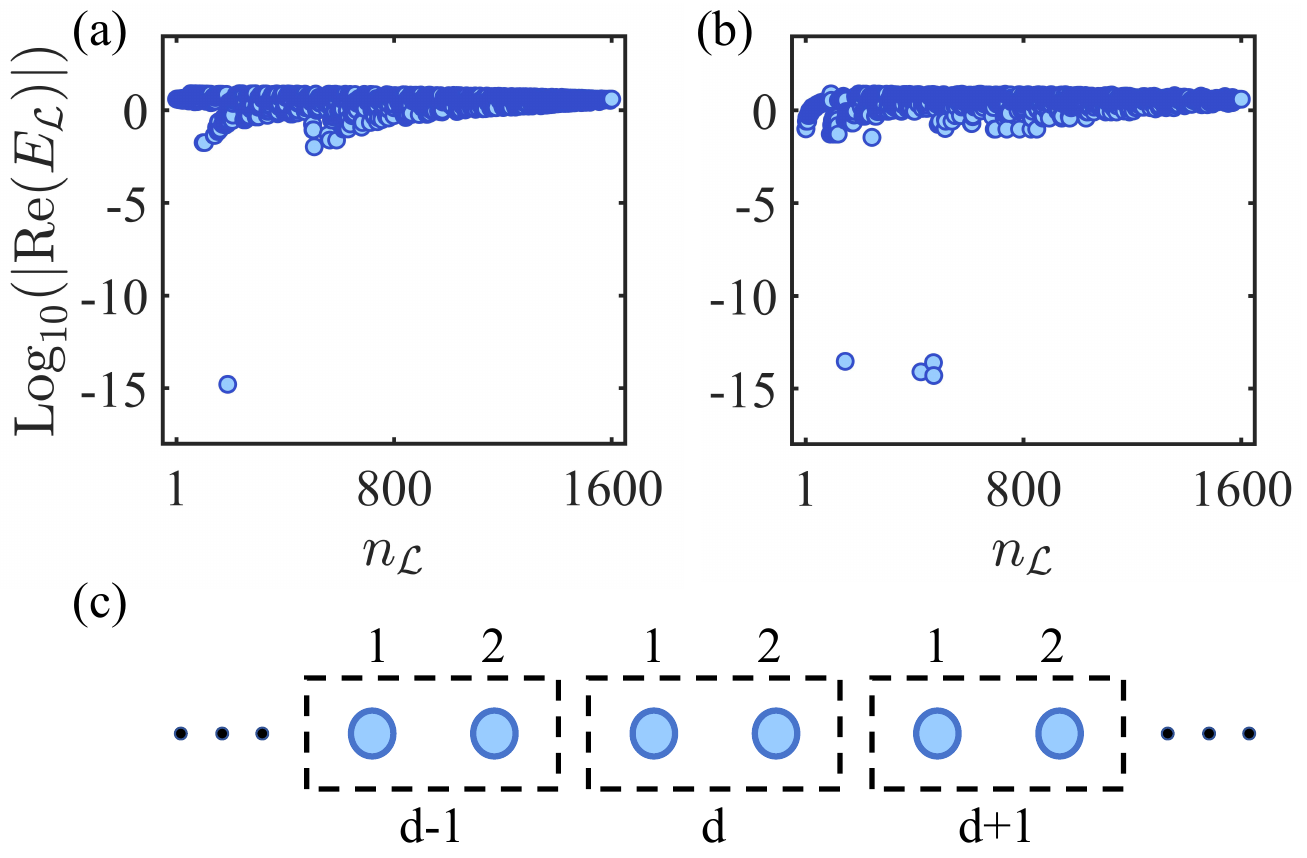}}
	\caption{The absolute value of the real parts of the eigenvalues of the Liouvillian superoperator for (a) $q=1$ and (b) 
		$q=2$. Here, we fix $a=1$ and $L=20$.
		(c) To explain the emergence of the fourfold steady state for $q=2$, we redefine each pair of lattice sites on a single chain of the cross-stitch model as a new unit.}
	\label{fig5}
\end{figure}

We take the cross-stitch model as an example to study the emergence of multiple steady states for $q>1$. When $q>1$, it is numerically straightforward to show that the steady state of the system still occupies the FB region, 
but the number of steady states will increase. Fig.~\ref{fig5}(a) and Fig.~\ref{fig5}(b) respectively display the real parts 
of the eigenvalues, $\text{Re}(E_{\mathcal{L}})$, of the Liouvillian superoperator $\mathcal{L}$ for $q=1$ and $q=2$ with fixed $a=1$. Here  $\mathcal{L}$ is represented as a $(2L)^2\times (2L)^2$ matrix, where $L$ denotes the number of unit cells.
One can see that for $q=1$ and $q=2$, there are numerically one and four values, respectively, for which $\text{Re}(E_{\mathcal{L}})\rightarrow 0$, indicating that the system has one and four steady states.
To understand why four steady states appear when $q=2$, we regroup the $q$ sites on each chain into a single unit, labeled as $d$, as shown in Fig.~\ref{fig5}(c). The bond dissipation with 
$q=2$ affects next-nearest-neighbor sites, meaning that sites $1$ and $2$ in different units are independently influenced. This leads to the formation of two dark states that satisfy Eq.(2) in the main text:
$|\Psi^{(1)}_{\text{dark}}\rangle = \sum_{d} A |\phi_{\text{CLS}}^{2d-1}\rangle$ and $|\Psi^{(2)}_{\text{dark}}\rangle = \sum_{d} B |\phi_{\text{CLS}}^{2d}\rangle$, both of which lie on the FB. The steady-state subspace spanned by these two dark states has dimension $2$, with the corresponding density matrix having dimension $4$. The steady-state density matrix is given by: $\rho_{ss}=\sum_{i,j=1}^2C_{ij}|\Psi^{(i)}_{\text{dark}}\rangle\langle\Psi^{(j)}_{\text{dark}}|$, where the coefficients 
$C_{ij}$ form a positive semi-definite, Hermitian and unit trace matrix. 
The method of utilizing such bond dissipation with $q>1$ to prepare multiple steady states can also be easily extended to systems without FBs, potentially leading to important applications.

\subsection*{V. Bond dissipation applied to 2D flat-band systems}

\subsubsection*{A. Twisted-bilayer square lattice}
\begin{figure}[h]
	\centering
	\includegraphics[width=0.6\linewidth]{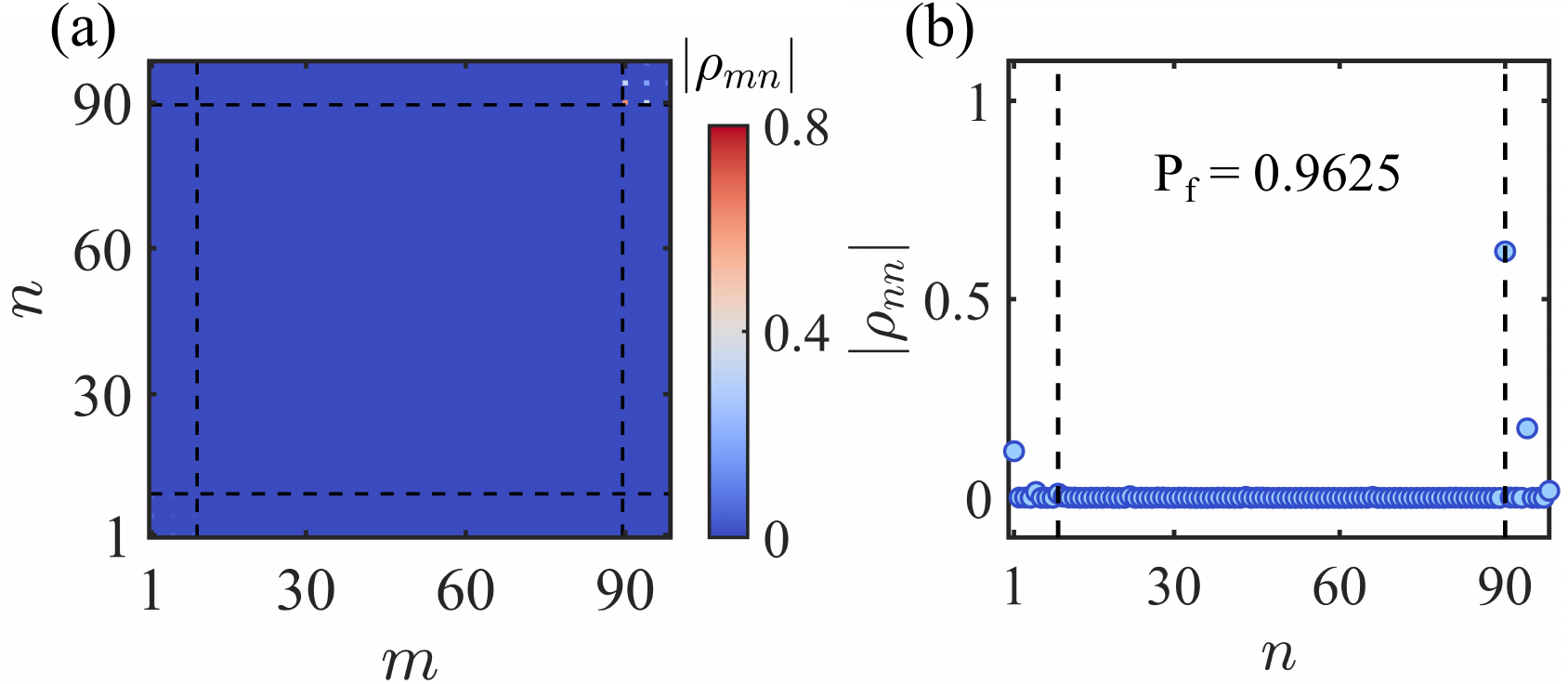}
	\caption{All parameters are the same as those used in Fig. 4 of the main text, except that the bond dissipation parameter is changed from $a=-1$ to $a=1$. The horizontal and vertical axes in panels (a) and (b) correspond to those in Fig. 4(c) and Fig. 4(d) of the main text, respectively.}
	\label{figS4}
\end{figure}

In the main text, we apply bond dissipation to a 2D twisted-bilayer square lattice, which hosts a nearly FB at an appropriate twist angle $\theta$. We find that the dissipation drives particle occupation from the dispersive band to the FB. Here, we modify the bond dissipation parameter from $a=-1$ to $a=1$, while keeping all other parameters unchanged from the main text. We find that the steady state remains predominantly occupied in the FB [Fig. \ref{figS4}(a)], with a FB occupation of $P_f=0.9625$ [Fig. \ref{figS4}(b)], despite the FB subspace accounting for only a small portion of the total Hilbert space. The physical roles of $a=-1$ and $a=1$ correspond to flipping in-phase configurations to out-of-phase ones and vice versa, both of which can drive particles from the dispersive band to the FB. This highlights the flexible tunability of the relative phases between localized wavefunctions in the FB.

\subsubsection*{B. 2D Lieb lattice}
\begin{figure}[h]
	\centering
	\includegraphics[width=0.9\linewidth]{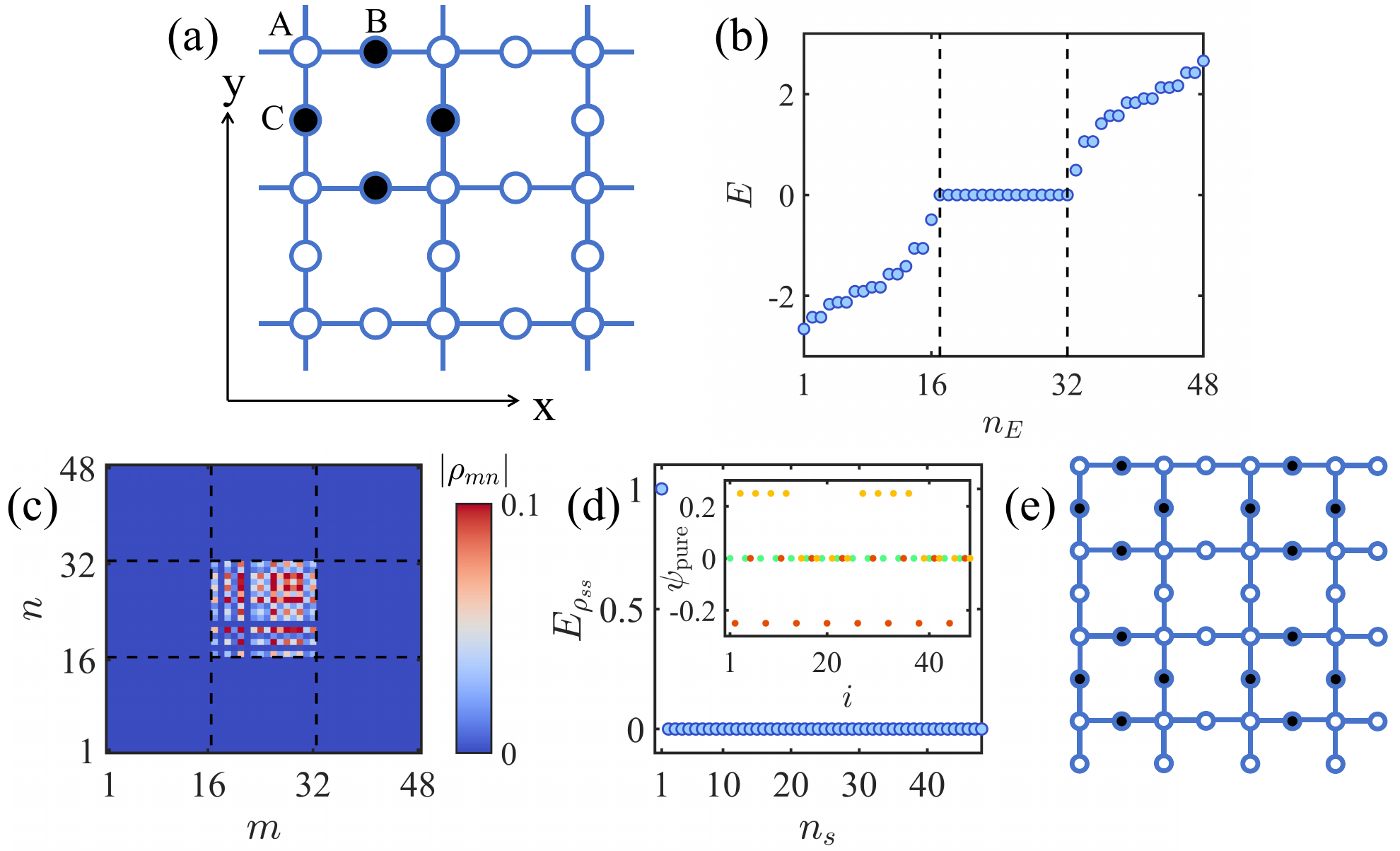}
	\caption{(a) Schematic of the 2D Lieb lattice. Black dots mark the sites occupied by the CLSs, which span two unit cells in both directions and are localized on sublattices $B$ and $C$. (b) Energy spectrum of the 2D Lieb lattice, showing a FB at 
		$E=0$. (c) Absolute values of the steady-state density matrix elements projected onto the eigenbasis of the Hamiltonian, indicating that the occupation is confined to the FB subspace.
		(d) Spectrum of the Liouvillian superoperator showing a single zero eigenvalue (i.e., a unique steady state). The inset shows the real-space distribution of this steady state. Green, red, and yellow dots correspond to sublattices $A$, $B$, and $C$, respectively.
		(e) Real-space distribution of the steady state over all lattice sites. Black dots highlight the locations with nonzero occupation, consistent with the CLS structure.}
	\label{figS5}
\end{figure}

In Section III, we analyzed the effect of bond dissipation on the 1D Lieb lattice. Here, we extend our investigation to the 2D Lieb lattice. The 2D Lieb lattice consists of three sublattice sites per unit cell, labeled $A$, $B$, and $C$, as shown in Fig.~\ref{figS5}(a). As in the 1D case, we set the hopping amplitude to $1$ for all intra-cell and inter-cell connections: within each unit cell between sites $A$-$B$ and $A$-$C$, and between neighboring unit cells--specifically, from $B$ to $A$ in the $x$-direction, and from $A$ to $C$ in the $y$-direction. We fix the system size to be four unit cells along both the $x$- and $y$-directions. The energy spectrum of this system is shown in Fig.~\ref{figS5}(b), revealing a flat band at energy $E=0$. The corresponding CLSs, marked as black dots in Fig.~\ref{figS5}(a), occupy only the $B$ and $C$ sublattice sites and span two unit cells in both the $x$- and $y$-directions.

Each direction of the 2D Lieb lattice can be viewed as a series of 1D Lieb lattices, where the FB states in each direction can be understood as a $U=2$ CLS configuration. The bond dissipation operator $O_{j}=\left( c_{j}^{\dag }+ac_{j+q}^{\dag }\right) \left( 
c_{j}-ac_{j+q}\right)$ can be applied along either the $x$- or $y$-direction, both yielding the same result. We fix $\kappa=1$, $\Gamma=1$, and $a=1$. Based on the discussion in the 1D case, when applied along the $x$-direction, we set $q=4$ for the chains containing sublattices $A$ and $B$, and 
$q=2$ for the chain containing sublattice $C$. When applied along the $y$-direction, we set 
$q=4$ for the chains containing sublattices $A$ and $C$, and $q=2$ for the chain containing sublattice $B$.
As shown in Fig.~\ref{figS5}(c), the steady state is fully confined to the FB subspace. Similar to the 1D case, there is a unique pure steady state [Fig.~\ref{figS5}(d)], which can be expressed as a linear combination of CLSs. The spatial structure of these CLSs is illustrated in Fig.~\ref{figS5}(e).

\subsubsection*{C. topological checkerboard lattice}
\begin{figure}[h]
	\centering
	\includegraphics[width=0.9\linewidth]{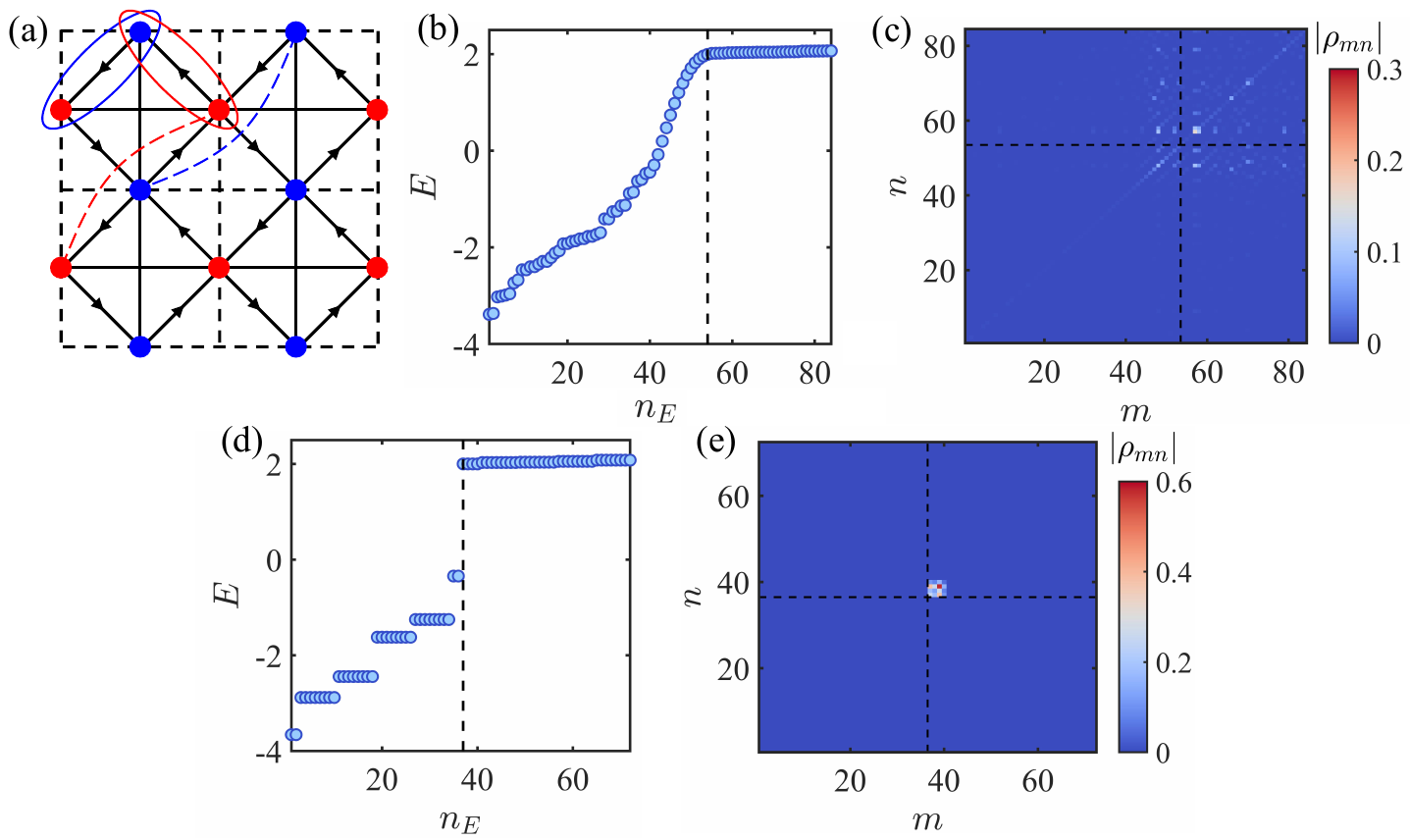}
	\caption{(a) Schematic of the topological checkerboard lattice, where arrows and solid (dashed) lines represent NN and NNN hoppings, respectively. The arrow direction indicates the sign of the phase factor in the NN hopping terms. Two representative NNNN hoppings are shown as dashed curves. The two solid circles indicate the two directions of bond dissipation. (b,d) Energy spectra of the model.
		(c,e) Absolute value of the steady-state density matrix elements.
		(b,c) are obtained with open boundary conditions and system size $L=84$, while (d,e) are obtained with periodic boundary conditions and system size $L=72$. The dashed lines in (b-e) indicate the boundaries between the dispersive band and the FB. The bond dissipation parameters are chosen as $q=1$ and $a=-1$.}
	\label{figTFB}
\end{figure}

Topological FBs have garnered significant research interest in recent years. To investigate the impact of bond dissipation on such systems, we examine a two-band model on a checkerboard lattice that hosts a topologically nontrivial nearly FB through the introduction of specific next-nearest-neighbor (NNN) and next-next-nearest-neighbor (NNNN) hoppings~\cite{KSun2011S}. Its Hamiltonian reads:
\begin{equation}
	H    = -J\sum_{\langle i,j\rangle}e^{i\phi_{ij}}\left(c_{i}^{\dagger}c_{j}+\text{H.c.}\right)  
	-\sum_{\langle\langle i,j\rangle\rangle}J_{ij}^{\prime}\left(c_{i}^{\dagger}c_{j}+\text{H.c.}\right) 
	-J^{\prime\prime}\sum_{\langle\langle\langle i,j\rangle\rangle\rangle}\left(c_{i}^{\dagger}c_{j}+\text{H.c.}\right),
\end{equation}
where $\langle i,j \rangle$, $\langle\langle i,j \rangle\rangle$, and $\langle\langle\langle i,j \rangle\rangle\rangle$ denote summations over nearest-neighbor (NN), NNN, and NNNN pairs, respectively. The NN hopping amplitude is set to $J = 1$. The phase factor 
in the NN hopping terms is given by
$\phi_{ij} = \pm\phi$, where the sign is determined by the orientation of the arrows in Fig.~\ref{figTFB}(a); we set $\phi = \pi/4$. The NNN hopping amplitude $J'_{ij}$ takes the value $J'_1$ ($J'_2$) when the two sites are connected by a solid (dashed) line in Fig.~\ref{figTFB}(a). The parameter values are selected as $J'_1 = -J'_2 = 1/(2+\sqrt{2})$ and $J'' = 1/(2+2\sqrt{2})$ for the NNNN hopping. Under these conditions~\cite{KSun2011}, the top band becomes nearly flat [Fig.~\ref{figTFB}(b)], and both bands acquire nonzero Chern numbers of $\pm 1$.

We now consider the effect of bond dissipation on this system. Due to the absence of a CLS in this nearly flatband system, the dark-state condition cannot be directly applied to engineer the bond dissipation. Instead, we design the bond dissipation along two directions, as shown in Fig.~\ref{figTFB}(a). By choosing the dissipation parameters 
$q=1$ and $a=-1$ to select out-of-phase states, we find that under open boundary conditions (OBC), the steady state predominantly occupies the nearly flatband portion of the spectrum [Fig.~\ref{figTFB}(c)], while additional occupation also appears on the boundary states. Since the steady state only occupies the nearly flatband and the boundary states, as shown in the main text, removing the bond dissipation does not change the distribution of the steady state over the eigenstates. The states within the flatband exhibit localized characteristics. Therefore, the inclusion of bond dissipation enables the quantized transport properties of the boundary states to manifest, while avoiding the influence of the bulk dynamical behavior. 
To eliminate the influence of these boundary states, we consider periodic boundary conditions, where the spectrum is shown in Fig.~\ref{figTFB}(d). Under bond dissipation, the steady-state occupation of the nearly flatband approaches unity [Fig.~\ref{figTFB}(e)].

\end{document}